\documentclass[english,aps,prl, twocolumn,superscriptaddress]{revtex4-2}
\usepackage[utf8]{inputenc}
\usepackage{amsmath}
\usepackage{amssymb}
\usepackage{bm}
\usepackage{graphicx}
\usepackage{hyperref}
\usepackage{braket}
\usepackage{babel}
\usepackage{mathtools}
\usepackage{xspace}
\usepackage{xcolor}
\usepackage{physics}
\usepackage{bbm}

\begin{document}
\title{Self-similar phase diagram of the Fibonacci-driven quantum Ising model}
\author{Harald Schmid}
\affiliation{\mbox{Dahlem Center for Complex Quantum Systems and Fachbereich Physik, Freie Universit\"at Berlin, 14195 Berlin, Germany}}
\author{Yang Peng}
\affiliation{\mbox{Department of Physics and Astronomy, California State University, Northridge, CA 91330, USA}}
\author{Gil Refael}
\affiliation{\mbox{Institute of Quantum Information and Matter and Department of Physics,} \mbox{ California Institute of Technology, Pasadena,
CA 91125, USA}}
\author{Felix von Oppen}
\affiliation{\mbox{Dahlem Center for Complex Quantum Systems and Fachbereich Physik, Freie Universit\"at Berlin, 14195 Berlin, Germany}}
\begin{abstract}
We study a stroboscopic quantum Ising model with Fibonacci dynamics. Focusing on boundary spin correlation functions in long but finite chains, our simulations as well as analytical arguments reveal a self-similar phase diagram exhibiting regions with Majorana zero modes (MZM) as well as Majorana golden-ratio modes (MGM). We identify the self-similarity transform which governs the evolution of the phase diagram with increasing simulation time. Integrability-breaking perturbations lead to a temporal decay of the boundary spin correlations, ultimately limiting the self-similarity of the phase diagram. Our predictions are testable with current quantum information processors.
\end{abstract}

\date{\today}
\maketitle
%

{\em Introduction.---}Quasiperiodicity has been a remarkably rich source of intriguing physics. In quantum systems, it is known to lead to 
self-similar spectra as exemplified by the Hofstadter butterfly for Bloch electrons in strong magnetic fields \cite{Harper1955,Hofstadter1976}. Moreover, quasiperiodic potentials are intermediate between periodic and disordered, which make them model systems in relation to Anderson localization \cite{Aubry1980}. In quantum dynamics, quasiperiodic driving introduces dynamical localization in the energy domain, which protects systems such as the paradigmatic kicked rotor \cite{Shepelyansky1983,Grempel1984,Moore1994,Blekher1992} against heating. Quasiperiodic driving with multiple incommensurate frequencies realizes dynamical localization transitions between phases with and without protection against heating in the universality class of the 3D Anderson transition \cite{Casati1989,Chabe2008}.

More recent research on quantum dynamics with quasiperiodic driving focuses on quantum many body systems. This is motivated by fundamental interest as well as implementations 
on quantum information processors. In particular, driven spin chains are readily implemented and raise numerous fascinating and challenging questions. These include the fate of 
many body localization and glassy dynamics in quasiperiodically driven systems \cite{Dumitrescu2018}, the possibility of  discrete-time quasicrystals \cite{Zhao2019}, as well as the heating behavior of critical systems \cite{Lapieree2020, Wen2021,Maity2019}. Recent work has also focused on the quantum dynamics of topological phases in quasiperiodically driven systems \cite{Martin2017,Peng2018,Yang2018,Else2020,Long2021,Friedman2022,Long2024,Qi2024}. 

Here, we introduce and study the concept of a scale-dependent dynamical phase diagram of quasiperiodically driven many body systems. While their long-time dynamics has been extensively studied, little attention has been paid to phase diagrams. We address this question for a quantum Ising model stroboscopically driven with a quasiperiodic Fibonacci drive. The quantum Ising model and its phase diagram are already well understood both in the Hamiltonian \cite{Lieb1961, Pfeuty1970} and the Floquet incarnations \cite{Jiang2011, Bauer2019, Mitra2019,  Schmid2024}, which made them central to Floquet time crystals \cite{Wilczek2012, Else2016, Khemani2016, Else2017, Krzysztof2017, Khemani2019, Else2020TC, Mi2021} and frequently implemented models on current quantum information processors \cite{Mi2021,Mi2022}.  

In our context, boundary modes in finite quantum Ising chains (related to Majorana modes in the fermionized version aka Kitaev chain) provide a convenient diagnostic of the phase. We show that these can be used to uncover and understand the scale-dependent and self-similar phase diagram of the Fibonacci-driven quantum Ising model. Moreover, the quantum Ising model readily admits introducing symmetry- and integrability breaking perturbations. We find that breaking of integrability leads to a decay of boundary correlations, providing an intrinsic cutoff of the self-similarity of the phase diagram. 

\begin{figure*}
    \centering
    \includegraphics[width=\linewidth]{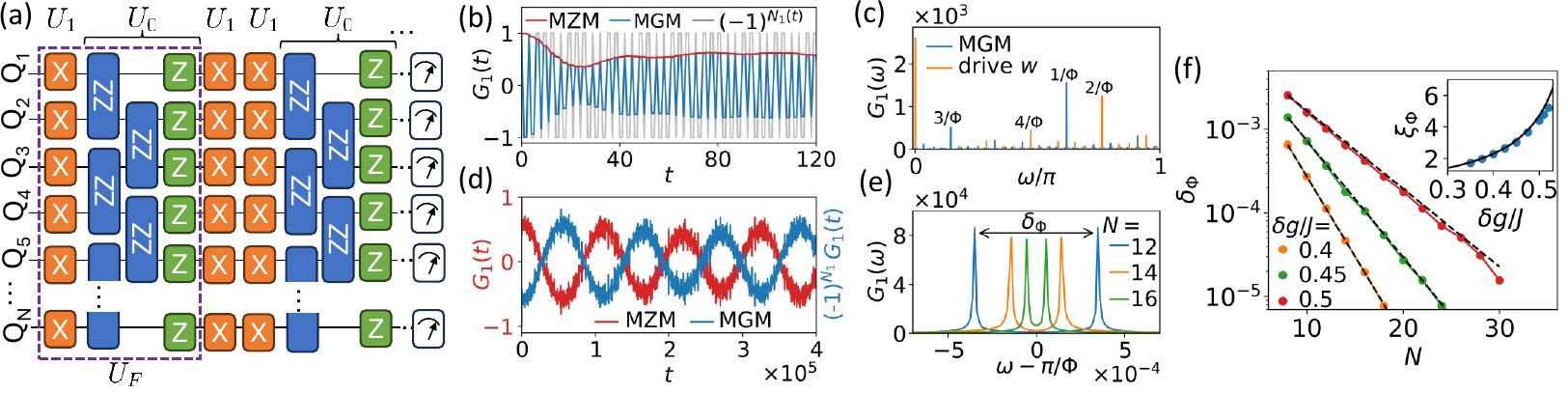}
    \caption{Quasiperiodically driven quantum Ising model. 
    (a) Quantum circuit for Fibonacci drive. Purple box: Floquet operator $U_F$ of Floquet quantum Ising model. (b) Boundary correlation function $G_1(t)=\langle Z_1(t)Z_1(0) \rangle$ without longitudinal fields in the MZM (red) and MGM (blue) regime (gray: ideal MGM case $g=1$). (c) Fourier transform $G_1(\omega)$ (MGM) peaks at $2\pi(n+1/2)/\Phi$ (blue), while driving sequence $w$ peaks at $2\pi n/\Phi$ (orange). (d) $G_1(t)$ for longer simulation times with oscillations due to Majorana hybridization (superimposed on fast quasiperiodic pattern for MGM). (e) Hybridization splitting of dominant MGM peak. (f) Splitting $\delta_\Phi$ decreases exponentially with system size $N$. Inset: Majorana localization length $\xi_\Phi$ extracted from splittings diverges at $J=\delta g \Phi$. Parameters: $J=0.1$ (all), $g=0.04$ (MZM), $g=0.96$ (MGM), (b,c) $N=100$, (d) $N=16$.}
    \label{fig:figure1}
\end{figure*}

{\em Fibonacci-driven quantum Ising model.---}Our driven quantum Ising model obeys the Fibonacci rule 
\cite{Dumitrescu2018, Friedman2022} 
\begin{align}
    U(f_{n-1})U(f_{n})=U(f_{n+1}),
\end{align}
with initial condition $U(f_0) \to U_0$ and $U(f_1) \to U_1$. Here, $U(f_n)$ denotes the time evolution operator at time $f_n$, with $f_n$ being the $n$th Fibonacci number ($f_0=f_1=1$). The basic unitaries  
\begin{align}
U_0=e^{\frac{i\pi J}{2}\sum^{N-1}_{j=1}Z_jZ_{j+1}},
    \quad
U_1=e^{\frac{i\pi g}{2}\sum^N_{j=1}X_j}
\label{eq: elementary unitaries}
\end{align}
respectively implement the two-qubit exchange and the single-qubit transverse-field gates on the qubits of the chain (Pauli operators $X_j$ and $Z_j$ at site $j$). As $U(f_n)$ has $f_n$ factors, we define $U(t)=\dots U_0U_1U_1U_0U_1$ for general (integer) times $t<f_n$ by retaining the $t$ rightmost factors in $U(f_n)$ [see quantum circuit in Fig.\ \ref{fig:figure1}a]. 

\begin{figure*}
    \centering
    \includegraphics[width=\linewidth]{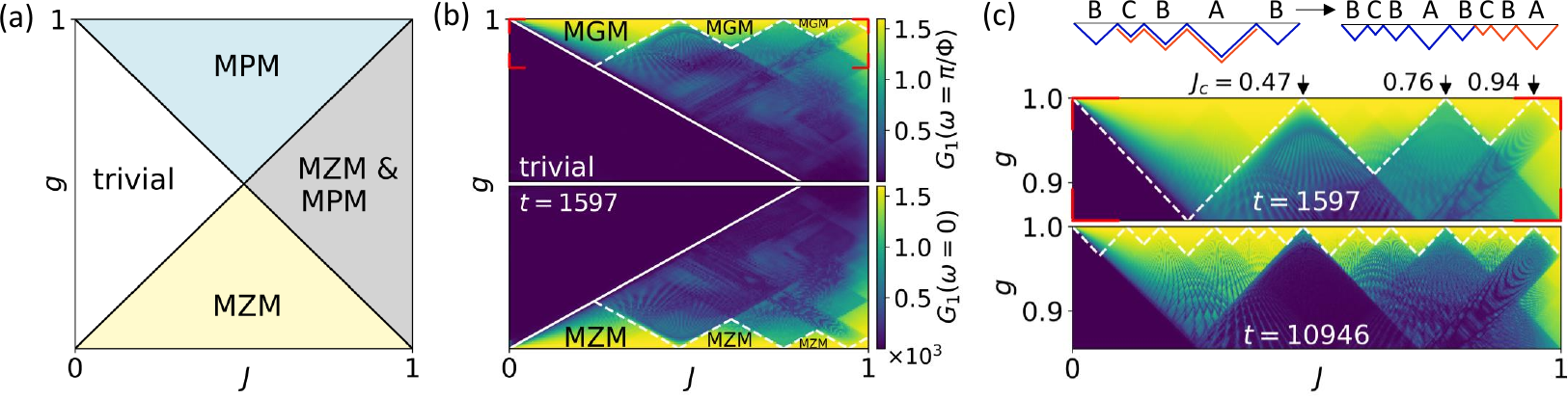}
    \caption{Phase diagrams of driven quantum Ising models. (a) Floquet model with MZM and MPM. (b) Quasiperiodic Fibonacci drive. Map of peak heights of $G_1(\omega)$  at $\omega=0$ (lower half) and $\omega=\pi/\Phi$ (upper half), indicating stable/unstable regions of MZM and MGM. White solid lines: dominant phase boundaries $J=g\Phi$ and $J=(1-g)\Phi$. Dashed cones originating from unstable points $J_c$ show additional phase boundaries for $t=1597$ time steps.  
    (c) Zoom of stable/unstable regions in the MGM phase for $t=1597$ and $t=10946$. Longer simulations reveal a self-similar structure with more instability points. Top: Self-similarity transformation describing the evolution of stable/unstable regions with simulation time $t$. 
    Parameter: $N=40$.}
    \label{fig:figure2}
\end{figure*}

We extract the phase diagram of the model from numerical results for the infinite-temperature correlation function $G_j(t)=\langle Z_j(t)Z_j(0)\rangle$, which averages over the entire many-body spectrum, $\langle\ldots\rangle = 2^{-N}\mathrm{tr}[\ldots]$. This is motivated by the Floquet quantum Ising model (time evolution operator $U(t)=[U_F]^t$ with Floquet operator $U_F=U_0U_1$ as illustrated in Fig.\ \ref{fig:figure1}a). Its phases (Fig.\ \ref{fig:figure2}a) are characterized by the appearance of Majorana zero modes (MZMs) and Majorana $\pi$ modes (MPMs) at the ends of a finite chain. This is reflected in characteristic behavior of the boundary correlation functions $G_1(t)$ and $G_N(t)$ \cite{Mitra2019,Mi2022,Schmid2024}. After an initial transient,
the boundary correlation function plateaus in the MZM case and exhibits period doubling $G_1(t) \propto (-1)^t$ in the MPM case. This is reflected in $\omega=0$ and $\omega =\pi$ peaks in the Fourier transform $G_1(\omega)$, respectively. Bulk spin correlations $G_j(t)$ ($j\neq 1,N$) decay rapidly.

Figure \ref{fig:figure2}b shows a phase diagram of the Fibonacci-driven quantum Ising model obtained in an analogous way. For transverse fields near $g=0$, we find that the boundary correlation functions exhibit plateau-like behavior and hence an $\omega=0$ peak similar to the Floquet quantum Ising model. Figure \ref{fig:figure2}b (lower half) maps the height of this peak in the $g-J$ plane, revealing triangular regions exhibiting MZMs. Cones with slopes of about $\pm 0.62$ originate from unstable  points at $g=0$, tracing critical lines. $G_1(\omega=0)$ drops sharply across these lines. 

Near the line $g=1$, the dominant Fourier peak of the boundary correlation function has frequency $\omega \simeq 0.62 \pi$ [Fig.\ \ref{fig:figure1}b,c]. This contrasts with the Floquet case, where the dominant peak appears at $\omega=\pi$. For reasons that will become clear below, we refer to the underlying boundary mode with frequency $\omega \simeq 0.62 \pi$ as the Majorana golden-ratio mode (MGM). We map the MGM peak in Fig. \ref{fig:figure2}b (upper half). As emphasized by the zoom in Fig.\ \ref{fig:figure2}c, the fine structure of the MGM regions is due to unstable points at $J_c \simeq 0.76, 0.47,\dots$ on the line $g=1$. The
fine structure around $g=1$ [Fig.\ \ref{fig:figure2}b,c] mirrors the one near $g=0$ for MZMs. 

Remarkably, we find that the fine structure depends on the simulation time, i.e., the time interval over which the Fourier transform $G_1(\omega)$ is computed. This is illustrated in Fig.\ \ref{fig:figure2}c, which shows phase diagrams obtained at two Fibonacci times $f_{16}=1597$ and $f_{20}=10946$. At larger times, the phase diagram exhibits more unstable points with their associated phase boundaries. We find by inspecting a sequence of phase diagrams for consecutive Fibonacci times that up to boundary effects, the structure of the stable triangular regions is generated by iterating the initial self-similarity transformation shown in \ref{fig:figure2}c (top). For each Fibonacci time, there are three triangle types (labeled A,B,C) with distinct base lengths (ratios $\simeq 0.62$), but identical slopes of the isosceles. In each step, there are $f_n$ triangles (starting with $f_4=5$), which one concatenates with a copy truncated by $f_{n-3}$ ($f_{n-4}$) triangles on the left (right).

{\em Quasiperiodic driving and boundary correlation functions.}---We turn to understanding this self-similar phase diagram. An explicit expression for the time-evolution operator of the Fibonacci-driven quantum Ising model is $U(t)=U_{w(t)}\ldots U_{w(1)}$ with driving pattern $w(t)=\lfloor (t+1)/\Phi\rfloor - \lfloor t/\Phi\rfloor$, where $\Phi=(1+\sqrt{5})/2$ is the golden ratio, see \cite{SI}. The number of $U_1$ factors in $U(t)$ equals $N_1(t) = \sum_{t'=1}^t w(t') = \lfloor (t+1)/\Phi\rfloor$. The Fourier transform of the driving pattern $w(t)$ has peaks at $2\pi n/\Phi$ ($n\in\mathbb{Z}$) [Fig.~\ref{fig:figure1}c]. We note that due to the stroboscopic time evolution, these frequencies can be folded back into the basic frequency interval $[-\pi,\pi]$. The irrational nature of $\Phi$ implies that the model involves quasiperiodic driving, with a dense spectrum of measure zero in $[-\pi,\pi]$.

A Jordan-Wigner transformation to Majorana operators, $a_{2j-1}=(\prod_{l<j}X_l)Z_j$ and $a_{2j}=(\prod_{l<j}X_l)Y_j$, maps the basic unitaries into the free-fermion unitaries 
\begin{align}
U_0=e^{\frac{\pi J}{2}\sum^{N-1}_{j=1}a_{2j}a_{2j+1}}, \quad
    U_1=e^{\frac{\pi g}{2}\sum^N_{j=1}a_{2j-1}a_{2j}}. 
\end{align} 
The transverse field couples Majoranas from the same site, the exchange coupling from different sites. The free-fermion nature implies that the time evolution of the Majorana operators, $a_j(t)=U^\dagger(t)a_jU(t)=\sum_l W_{jl}(t)a_l$, can be described in terms of the Bogoliubov-de Gennes time evolution operator $W(t)$. This allows for efficient computation of boundary spin correlations as $G_1(t)=\expval{ a_1(t)a_1(0)} = W_{11}(t)$ \cite{SI}.

We gain intuition for the boundary correlation function by considering special cases. First consider $J=0$, where $U_0=\mathbbm{1}$. Then, eigenstates of $U(t)=U_1^{N_1(t)}$ are eigenstates of $U_1$, which dimerizes Majoranas of the same site. One readily finds $G_j(t)=\cos(\pi g N_1(t))$ for all $j$. Its Fourier transform $G_j(\omega)$ has a dense spectrum of peaks at frequencies $(\pm\pi g + 2\pi n)/ \Phi$. The peak replicas reflect the incommensurate frequencies $2\pi n/\Phi$ of the  quasiperiodic driving pattern $w(t)$. Consistent with the phase diagram, there are no stable boundary modes along the line $J=0$.

Next, consider $g=0$, where $U_1=\mathbbm{1}$. Thus, eigenstates of $U(t)=U_0^{t-N_1(t)}$ are eigenstates of $U_0$, which dimerizes Majoranas of neighboring sites. One finds $G_j(t)=\cos(\pi J (t-N_1(t)))$ with a dense Fourier spectrum for all bulk sites ($j\neq 1,N$). In contrast, the boundary correlation function becomes $G_1(t)=1$, since the two end Majoranas decouple and turn into Majorana zero modes (MZMs) \cite{SI}. Unlike the finite-frequency peaks of $G_j(\omega)$ for bulk sites, the $\omega=0$ MZM peak of $G_1(\omega)$ does not have replicas due to the quasiperiodicity.

Finally, for $g=1$, the transverse-field unitary flips all spins, $U_1=P=\prod_j X_j$, so that both $U_0$ and $U_1$ are nontrivial. Due to the decoupling of the end Majoranas and the identification $a_1=Z_1$, we find the boundary correlation function $G_1(t)=(-1)^{N_1(t)}$, while the bulk dynamics depends on the value of $J$ \cite{SI}. Unlike for $g=0$, $G_1(\omega)$ exhibits a dense spectrum after backfolding,
with peaks at the half-integer harmonics $2\pi(n+1/2)/\Phi$ of the golden ratio. This motivates the designation as MGM. Importantly, the MGM spectrum differs from the driving pattern $w(t)$ with peaks at $2\pi n/\Phi$ [Fig.\ \ref{fig:figure1}c].

The dominant phase boundaries [full lines in Fig.\ \ref{fig:figure2}b] can be understood by writing $U(t) \simeq e^{-iH^\mathrm{MZM}_\mathrm{eff}t}$ ($g$ small) and $U(t)\simeq P^{N_1(t)} e^{-iH^\mathrm{MGM}_\mathrm{eff}t}$ ($g\approx 1$). For $g$ and $J$ small, higher commutators can be neglected, and the effective Hamiltonian becomes the quantum Ising model,
\begin{align}
    H^\mathrm{MZM}_\mathrm{eff} \simeq -\frac{ \pi g }{2\Phi}\sum_j X_j-\frac{ \pi J }{2\Phi^2}\sum_j Z_jZ_{j+1}.
    \label{eq: effective Hamiltonian}
\end{align}
$H^\mathrm{MGM}_\mathrm{eff}$ is obtained by replacing $g\to 1-g$ in $H^\mathrm{MZM}_\mathrm{eff}$. The $\Phi$-dependence of the prefactors reflects the multiplicities of $U_0$ and $U_1$ in $U(t)$. The dominant lines correspond to the self-duality lines $J=g\Phi$ (MZM) and $J=(1-g)\Phi$ (MGM) of the effective quantum Ising Hamiltonian. 

\begin{figure*}
    \centering    \includegraphics[width=\linewidth]{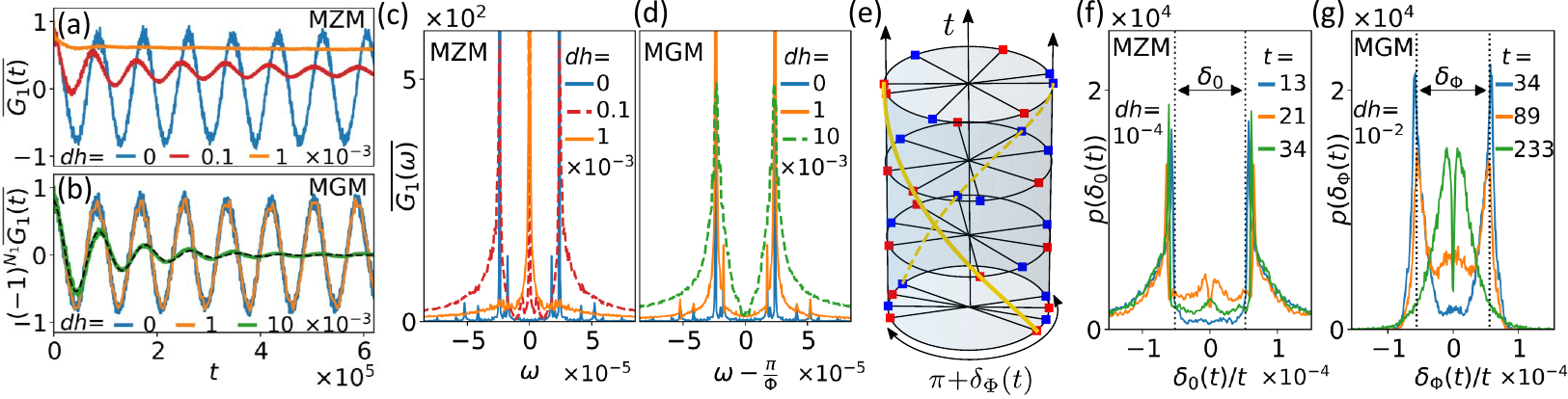}
    \caption{Boundary correlations with random longitudinal fields [full circuit in Fig.\ \ref{fig:figure1}(a)]. (a) MZM: 
    Longitudinal fields rapidly suppress 
    hybridization-induced oscillations in disorder-averaged $G_1(t)$ and polarize edge spins. (b) MGM: Oscillations remain robust up to much larger $dh$. Dashed line: fit to exponential decay. (c),(d) Fourier transforms $G_1(\omega)$. (e) For MGMs, many-body eigenphases of $U(t)$ are $\pi$-paired when $e^{i\pi N_1(t)}=-1$, up to a state-dependent splitting $\delta_\Phi(t)$ which grows linearly with $t$. (f)  Splitting distribution of zero pairs, normalized to splitting for $dh=0$ and  rescaled by $t$. Longitudinal fields ($dh=10^{-4}$) rapidly broaden the distribution. (g) Distribution of normalized splittings of $\pi$ pairs. Splittings mostly decrease for nonzero  $dh=10^{-2}$. Parameters: $N$=$10$, $\mathcal{N}$=$100$ realizations, $g$=$0.03$ (MZM), $g$=$0.97$ (MGM), $J$=$0.1$ (both).}
    \label{fig:figure3}
\end{figure*}

The fine structure of the phase diagram can be understood from the dynamics of the single-particle bulk modes. At the unstable points, the bulk gap collapses and the edge modes are suppressed. This can be seen from the single-particle time-evolution operator $W(t)$. The bulk dynamics is captured by a chain of $N$ sites with periodic boundary conditions, so that modes can be labeled by momentum $k$. In the Majorana basis, the elementary single-particle unitaries read \cite{SI}
\begin{align}
W_0(k) = e^{i\frac{k}{2}\sigma^z}e^{i\pi J\sigma^y}e^{-i\frac{k}{2}\sigma^z}, \quad 
W_1(k) =e^{-i\pi g\sigma^y}.
\end{align}
At $g=0$, we find $W(k,t)=[W_0(k)]^{\lfloor t/\Phi^2\rfloor}\approx e^{i\pi J (t/\Phi^2) \sigma_y}$. Here, the last step is accurate when $t$ is a Fibonacci time \cite{SI}. The unstable points can now be identified as $J_c = 2m\Phi^2$ with $m\in \mathbb{Z}$. At these points, we have $W(k,t) \approx \mathbbm{1}$, corresponding to a gapless bulk single-particle spectrum for all $k$. At small but nonzero $g$, the gap closes only at $k=0$, where $W(k=0,t) \approx e^{i\pi [J (t/\Phi^2) - g (t/\Phi)  ]\sigma_y}$. In view of the symmetries of the phase diagram under $J\to J\pm 2$ and $J\to -J$, this reproduces phase boundaries at
\begin{align}
    J \mp g  \Phi  = 2 m \Phi^2 \mod 2,
    \label{eq: critical J}
\end{align}
i.e., the dashed lines in Fig.\ \ref{fig:figure2}b.  This uses that for $1 < J < 2$, the symmetries imply a phase boundary at $2-J$. For MGM modes near $g=1$, one replaces $g \rightarrow 1-g$. This follows by analogous arguments,  using that the gap now closes at $k=\pi$ \cite{SI}. Equation \eqref{eq: critical J} implies that all phase boundaries have slopes of the same magnitude $1/\Phi$. 

All unstable points have irrational $J_{c}$, so that they eventually cover the entire lines $g=0,1$. To estimate the number of unstable points at a given simulation time $t$, we note that at Fibonacci times,  $t/\Phi^2$ approximates $\lfloor t/\Phi^2\rfloor$ with a precision which increases $\sim 1/t$. This implies that $W(k=0,t)\simeq 1$ provided that $\pi J_c/(\sqrt{5}t) \ll 1$, i.e., for $t\gg 2\pi |m|\Phi^2/\sqrt{5}$ \cite{SI}. Thus, the number of unstable points which are resolved increases linearly with simulation time $t$, consistent with our numerical results. 

{\em Majorana modes}.---The interpretation in terms of Majorana modes is supported by the long-time dynamics. For long simulation times, the boundary correlation functions display sinusoidal oscillations [Fig.\ \ref{fig:figure1}d]. For MGMs, these oscillations are superimposed on the quasiperiodic pattern $(-1)^{N_1(t)}$. The oscillations translate to splittings $\delta_{0,\Phi}$ of the Fourier peaks in $G_1(\omega)$ [Fig.\ \ref{fig:figure1}e]. The splitting can be interpreted as hybridization of the Majorana modes at the two ends. Indeed, we find an exponential dependence of the splittings $\delta_{0,\Phi}\sim e^{-N/\xi_{0,\Phi}}$ on $N$. The localization lengths $\xi_{0}$ ($\xi_\Phi$) of the MZM (MGM) mode extracted from fits of the splittings   diverge at the phase transition lines $J=g\Phi$ (MZM) and $J=(1-g)\Phi$ (MGM) [Fig.\ \ref{fig:figure1}f] and are consistent with the Majorana localization lengths $\xi_{0} \simeq - 1/\ln(g\Phi/J)$ ($\xi_{\Phi}\simeq - 1/\ln((1-g)\Phi/J)$) predicted by the effective Hamiltonian in Eq.\ (\ref{eq: effective Hamiltonian}) [inset Fig.\ \ref{fig:figure1}f].

The presence of MGM can be viewed as a generalization of strong modes \cite{Fendley2012, Jermyn2014,  Fendley2016, Sreejith2016,
Else2017,Kemp2017,Mitra2019,Yeh2023} to quasiperiodic systems. The instantaneous many-body spectrum of the circuit, $U(t)\ket{n(t)}=e^{-iE_n(t)}\ket{n(t)}$ [Fig.\  \ref{fig:figure3}e], has paired  eigenphases $E_n(t) \in [-\pi,\pi]$. The pairs are degenerate for even and $\pi$-paired for odd $N_1(t)$, with the pairing being strongest at Fibonacci times. This implies that the corresponding mode operator $\Psi_\Phi$ satisfies $\Psi_\Phi U(t) \simeq e^{i\pi N_1(t)} U(t)\Psi_\Phi$. 

{\em Random longitudinal field.---}We investigate the robustness of the self-similar phase diagram against breaking of integrability and  spin-flip symmetry $P$ by considering quenched random longitudinal fields. This
modifies the exchange gates
$U_1\rightarrow e^{\frac{i\pi J}{2}\sum_j Z_jZ_{j+1}} e^{\frac{i\pi}{2}\sum_j h_j Z_j}$ with  spatially disordered $h_j \in [-dh,dh]$. In computing $G_1(t)$, we replace the trace by sampling over random initial product states in the $\{Z_j\}$-basis.

Breaking of integrability introduces an intrinsic cutoff of the self-similar structure of the phase diagram. Due to the dense spectra, it generically opens scattering channels, which induce a decay of the boundary spin correlations. This effectively provides an upper limit for the simulation time and hence the structural details of the phase diagram. Figure \ref{fig:figure3} shows simulation results for the disorder-averaged boundary correlation function. In the MZM case, the amplitude of the hybridization-induced oscillations decreases with increasing disorder strength $dh$ and crosses over into a slowly decaying nonoscillatory dependence, see Fig.\ \ref{fig:figure3}a. In the MGM case, the envelope oscillations of the boundary spin correlation persist to larger disorder strength, but decay exponentially at long times [Fig.\ \ref{fig:figure3}b]. Corresponding Fourier transforms $G_1(\omega$) are shown in Figs.\   \ref{fig:figure3}c and d. 

Breaking of the spin-flip symmetry $P$ by the random longitudinal field allows for asymmetry between the MZM and MGM regions. A pronounced asymmetry is seen in the dependence of the distribution of zero- and $\pi$-splittings on the strength of the longitudinal field, see Fig.\ \ref{fig:figure3}f and g. This reflects the distinct response of degenerate and $\pi$-paired many body states to the longitudinal field \cite{Schmid2024}. The longitudinal field couples the two (zero or $\pi$) paired many-body eigenstates. This induces a splitting for degenerate levels, while the  coupling between $\pi$-paired levels pushes them closer to $\pi$. 

\textit{Conclusions.---}We find that quasiperiodically driven quantum many body systems exhibit dynamical phase diagrams with self-similar scale dependence.
While generically requiring protection by integrability, some features such as the Majorana-like golden-ratio modes which we find exhibit remarkable resilience against integrability-breaking disorder. Interestingly, self-similar phase diagrams should be experimentally accessible on current quantum processors \cite{SI}. We expect extended quasiperiodically driven models to exhibit more elaborate fractal phase diagrams. Investigating the interplay of the fractal phase diagrams and the long-term quantum dynamics would be particularly interesting in such models. 

\begin{acknowledgments}
\textit{Acknowledgments.---}We thank Kang Yang for useful discussions. Research at Freie Universit\"{a}t Berlin was
supported by Deutsche Forschungsgemeinschaft through CRC 183 as well as by the Einstein Research Unit on Quantum Devices. Work at CSUN and Caltech was supported by NSF PREP Grant No.\ PHY-2216774. G.R.\ is grateful for support through AFOSR MURI Grant No.\ FA9550-22-1-0339, as well as the Simons Foundation
and the Institute of Quantum Information and Matter,
an NSF Frontier Center. We thank the HPC service of ZEDAT, Freie Universität Berlin, for computing time \cite{Bennett2020}.
\end{acknowledgments}

\vfill



%

\onecolumngrid

\clearpage

\setcounter{figure}{0}
\setcounter{section}{0}
\setcounter{equation}{0}
\renewcommand{\theequation}{S\arabic{equation}}
\renewcommand{\thefigure}{S\arabic{figure}}

\onecolumngrid

\centerline{\large Supplementary Material:} 

\centerline{\large Self-similar phase diagram of the Fibonacci-driven quantum Ising model}

\section{Fibonacci words}
For convenience, we briefly review Fibonacci words, on which the Fibonacci drive is based. 
Fibonacci words are a quasiperiodic sequence over a binary alphabet ${0,1}$, generated by recursively  applying the following substitution rule 
\begin{equation}
0 \to 1, \quad 1 \to  10.
\end{equation}
Starting from the word $w_1 = 1$, one obtains $w_2 = 10$, $w_3 =101$, and $w_4 = 10110$, etc. The fixed point of this substitution rule is the infinite Fibonacci word $w = 10110101\dots$. The length of $w_i$ is the $i$th Fibonacci number  $f_i$, and  $w_i$ represents the first $f_i$ letters of the infinite word $w$.  Moreover, for $i\geq 3$, $w_i$ can be obtained by concatenating $w_{i-1}$ and $w_{i-2}$, denoted as $w_{i} 
 = w_{i-1} w_{i-2}$.

\begin{figure*}[b!]
    \centering
    \includegraphics[width=0.4\linewidth]{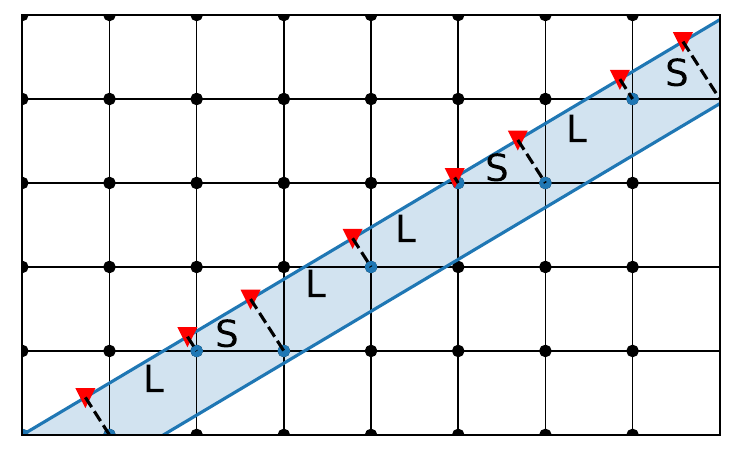}
    \caption{Projection method. A stripe is defined in a grid with unit spacing by two horizontally shifted lines with slope $\Phi^{-1}$. The projection of the grid points within the stripe (blue) onto the upper line defines a train of segments of two lengths $L$ (long) and $S$ (short). The order of the lengths corresponds to the quasi-periodic sequence Eq.\ \eqref{eq:w_word}.}
    \label{fig:projection_method}
\end{figure*}

Denoting the $n$th letter of the infinite Fibonacci word $w$ as $w(n)$, we can write explicitly
\begin{equation}
     w(n) = \lfloor(n+1)/\Phi\rfloor-\lfloor n/\Phi\rfloor=\begin{cases}
1 & {\rm if}\ (n/\Phi-\lfloor n/\Phi\rfloor)\in[1-\Phi^{-1},1)\\
0 & {\rm otherwise}
\end{cases}, \quad \Phi  = \frac{1+\sqrt{5}}{2},
\label{eq:w_word}
\end{equation}
which we will prove below. The number of ones and zeros in the word $w_t$ is 
\begin{align}
    N_1(t)=\lfloor t/\Phi \rfloor, 
    \quad 
    N_0(t)=t-\lfloor t/\Phi \rfloor, \quad N_0(t)+N_1(t)=t.
\label{eq:numberU1U0}
\end{align}

Equations \eqref{eq:w_word},\eqref{eq:numberU1U0} become intuitive by a geometric projection method \cite{Levine1986,Socoloar1986,Dumitrescu2018}, see Fig.\ \ref{fig:projection_method}. A two-dimensional square grid with unit spacing is cut by two straight lines $y_1=x/\Phi$ and $y_0=x/\Phi-1$. The lines define a stripe in which there are grid points $Q_n=\big(n,\lfloor n/\Phi \rfloor \big)$ (blue).  Projecting the grid points on the upper line $y_1$  defines points $P_n=(n+\lfloor n/\Phi \rfloor/\Phi )\big(1,\Phi^{-1}\big)$  (red triangles). The distance to the origin of the $n$-th point is
\begin{align}
    l_m=\frac{\sqrt{1+\Phi^2}}{\Phi^2}(m\Phi+\lfloor m/\Phi \rfloor).
\end{align}
The points $P_n$ divide $y_1$ into two types of segments $\Delta l_m = l_{m+1}-l_m$ of lengths $L=\sqrt{1+\Phi^2}$ and $S=\sqrt{1+\Phi^2}/\Phi$  with length ratio $L/S=\Phi$. The segments appear in the order described by \eqref{eq:w_word} and hence represent the $0$'s and $1$'s in the Fibonacci word. Note that the starting segment is $\Delta l_1=L$. The number of $L$'s ($1$'s) is $\lfloor n/\Phi \rfloor$. 

The explicit formula \eqref{eq:w_word} for  $w(n)$  is equivalent to the concatenation rule $w_i=w_{i-1}w_{i-2}$. The rule implies  
\begin{align}
    \lfloor(n+f_m+1)/\Phi\rfloor-\lfloor (n+f_m)/\Phi \rfloor=\lfloor(n+1)/\Phi\rfloor-\lfloor n/\Phi\rfloor, \quad 
    n\leq f_{m-1},
\end{align}
i.e., in a  word of length $f_{n+1}$ the sequence $n \leq f_{m-1}$ repeats for $f_m<n \leq f_{m+1}$. We now use the explicit form of the Fibonacci numbers $f_n=(\Phi^{n+1}-\Psi^{n+1})/(\Phi-\Psi)$ with $\Psi=-1/\Phi$ (we defined $f_0=f_1=1$). The ratio of two consecutive Fibonacci numbers approaches the golden ratio  exponentially with increasing $n$, 
\begin{align}
    \frac{f_n}{\Phi}=f_{n-1}+(-1)^{n}\Phi^{-(n+1)}.
\end{align}
We need to estimate whether the error remains sufficiently small such that $\lfloor n/\Phi\rfloor=\lfloor n/\Phi+(-1)^{m}\Phi^{-(m+1)}\rfloor$ for $n\leq f_{m-1}$. The term $n/\Phi$ is closest to an integer if $n=f_{k}$ is a Fibonacci number. The restriction $f_k\leq f_{m-1}$ keeps this error bounded which proves Eq.\ \eqref{eq:w_word}.

\section{Bogoliubov-de Gennes time evolution}
This section reviews the derivation of the Bogoliubov-de Gennes (BdG) time evolution operator $W(t)$ after Jordan-Wigner transformation (see main text). The time evolution is composed from
the elementary building blocks
\begin{align}
U_0=e^{\frac{i\pi J}{2}\sum^{N-1}_{j=1}Z_jZ_{j+1}}=e^{\frac{\pi J}{2}\sum^{N-1}_{j=1}a_{2j}a_{2j+1}},
    \quad
U_1=e^{\frac{i\pi g}{2}\sum^N_{j=1}X_j}=e^{\frac{\pi g}{2}\sum^N_{j=1}a_{2j-1}a_{2j}}.
\label{eq: elementary unitaries}
\end{align}
Due to the bilinear form of the unitaries in the Majorana representation,
the dynamics of the Majorana operators
\begin{equation}
a_j (t) = U^\dagger(t) a_j U(t)a_k=\sum_{l}W_{jl}(t)a_l
\label{eq:single-particle}.
\end{equation}
is of single-particle nature. The BdG time-evolution operator $W(t)$ of matrix size $2N\times 2N$ can be composed from the elementary building blocks
\begin{equation}
    U_0^\dagger a_{2i-1} U_0 = 
    \begin{cases}
        a_1, & i = 1 \\
        a_{2i-1}\cos \pi J - a_{2(i-1)}\sin\pi J, & i\neq 1
    \end{cases}, \quad
    U_0^\dagger a_{2i} U_0 = 
    \begin{cases}
        a_{2N}, & i = N \\
        a_{2i} \cos \pi J + a_{2i+1} \sin \pi J, & i \neq N. 
    \end{cases}
\end{equation}
and
\begin{equation}
U_1^\dagger a_{2i-1} U_1 = a_{2i -1}\cos\pi g + a_{2i}\sin \pi g, \quad 
U_1^\dagger a_{2i} U_1 = a_{2i}\cos \pi g - a_{2i - 1}\sin \pi g.
\end{equation}

The unitary transformations  $U_{i}$ ($i=0,1$) transform the Majorana operators linearly. Ordering the Majorana operators as $\boldsymbol{a} = (a_1,a_3,\dots, a_{2N-1}, a_2, a_4, \dots, a_{2N})^T$, the linear transformation becomes $U_i^\dagger \boldsymbol{a} U_i = W_i \boldsymbol{a}$, where $W_i$ has the explicit form
\begin{align}
W_0 &= \begin{pmatrix}
1\\
 & \cos\pi J\mathbb{I}_{N-1} & \sin\pi J\mathbb{I}_{N-1}\\
 & -\sin\pi J\mathbb{I}_{N-1} & \cos\pi J\mathbb{I}_{N-1}\\
 &  &  & 1
\end{pmatrix} = \mathbb{I}_2 \oplus e^{i \pi J \tilde{\sigma}_y} \otimes \mathbb{I}_{N-1}, \label{eq:W0} \\
W_1 &= 
\begin{pmatrix}
\cos\pi g\mathbb{I}_{N} & -\sin\pi g\mathbb{I}_{N}\\
\sin\pi g\mathbb{I}_{N} & \cos\pi g\mathbb{I}_{N}
\end{pmatrix} = e^{-i \pi g \sigma_y} \otimes \mathbb{I}_{N}. \label{eq:W1}
\end{align}
Here, $\mathbb{I}_2$ is the 2-by-2 identity in the basis of $(a_1, a_{2N})$, 
$\tilde{\sigma}_y$ is the $y$-Pauli matrix in the basis of $(a_{2j+1}, a_{2j})$ with $(N-1)$-by-$(N-1)$ identity $\mathbb{I}_{N-1}$ for indices $j=1,\dots, N-1$, and $\sigma_y$ is the $y$-Pauli matrix in the basis of $(a_{2j-1}, a_{2j})$ with $N$-by-$N$ identity $\mathbb{I}_{N}$ for indices $j=1,\dots,N$.

Using that $a_1=Z_1$, we see that $G_1(t)=W_{11}(t)$. We readily find from Eqs.\ (\ref{eq:W0}) and (\ref{eq:W1}) that $G_1(t)=1$ for $g=0$ and $G_1(t)=(-1)^{N_1(t)}$ for $g=1$, as quoted in the main text. 

For periodic boundary conditions, we can write 
\begin{equation}
a_{2j-1} = \frac{1}{N}\sum_{k} e^{i k j} a(k), \quad
a_{2j} = \frac{1}{N}\sum_{k} e^{i k j} b(k) 
\end{equation}
with Bloch momentum $k = 2\pi m/N$, $m = 0,1,\dots, N-1$. For a given $k$, the dynamics of the operators $a(k)$ and $b(k)$ under $U_0$ and $U_1$ are described by the 2-by-2 BdG time-evolution operators
\begin{align}
&W_0(k) = 
\begin{pmatrix}
\cos \pi J  & e^{ik}\sin \pi J \\
-e^{-ik} \sin \pi J &\cos \pi J
\end{pmatrix}= 
e^{i\frac{k}{2}\sigma^z}e^{i\pi J\sigma^y}e^{-i\frac{k}{2}\sigma^z},\\
&W_1(k) = 
\begin{pmatrix}
\cos \pi g  & -\sin \pi g \\
\sin \pi g &\cos \pi g
\end{pmatrix}
=e^{-i\pi g\sigma^y},
\end{align}
which act on the basis $(a(k), b(k))$. 

We identify dynamical gap closings in the case of MGM modes. For $g=1$, the BdG time evolution operator reads
\begin{align}
W(k,t)=(-1)^{N_1(t)}[W_0(k)]^{N_0(t)}
=(-1)^{N_1(t)}e^{i\frac{k}{2} \sigma^z} e^{i\pi J \sigma^y \lfloor t/\Phi^2 \rfloor } e^{-i\frac{k}{2}  \sigma^z}.
\label{eq: W g=1}
\end{align}
We see that at the critical points $J_c=2m \Phi^2$, this bulk time evolution operator at Fibonacci times $t=f_n$ reduces to the boundary time evolution operator deduced above and in the main text, up to an overall unitary transformation. This shows that the bulk gap closes for all $k$, mirroring the situation near  $g=0$. 

Away from $g=1$, the gap closes at the inversion-symmetric momentum $k=\pi$, where 
\begin{align}   W(k=\pi,t)=[W_0(k=\pi)]^{N_0(t)}[W_1(k=\pi)]^{N_1(t)} =  (-1)^{N_1(t)} e^{i\pi \delta g \lfloor t/\Phi \rfloor \sigma^y}e^{-i\pi J  \lfloor  t/\Phi^2 \rfloor \sigma^y}
\end{align}
with $\delta g=1-g$. The gap closing occurs at $J =2m\Phi^2+\Phi \delta g$, again mirroring the sitation near $g=0$.

In these considerations, we used that up to higher-order corrections in $1/f_n$, we have
\begin{equation}
    \lfloor \frac{f_n}{\Phi^2} \rfloor= \frac{f_n}{\Phi^2} + \frac{(-1)^n}{\sqrt{5}}\frac{1}{f_n}
\end{equation}
for $n$ odd. This directly follows from the identity $f_n=(\Phi^{n+1}-\Psi^{n+1})/\sqrt{5}$ and the observation that $\lfloor \frac{f_n}{\Phi^2}\rfloor = f_{n-2}$ with $n$ odd. The analogous equation for $n$ even takes the form 
\begin{equation}
    \lfloor \frac{f_n}{\Phi^2}\rfloor +1 = \frac{f_n}{\Phi^2} + \frac{(-1)^n}{\sqrt{5}}\frac{1}{f_n}.
\end{equation}
The accuracy of the gap closing is controlled by the second term on the right hand sides. The gap closing is well developed at Fibonacci time $t=f_n$ provided that $\pi J_c/(\sqrt{5}t) \ll 1$, i.e., for times  
\begin{equation}
t \gg 2\pi m \frac{\Phi^2}{\sqrt{5}}. 
\end{equation}

 \begin{figure*}
    \centering
\includegraphics[width=\linewidth]{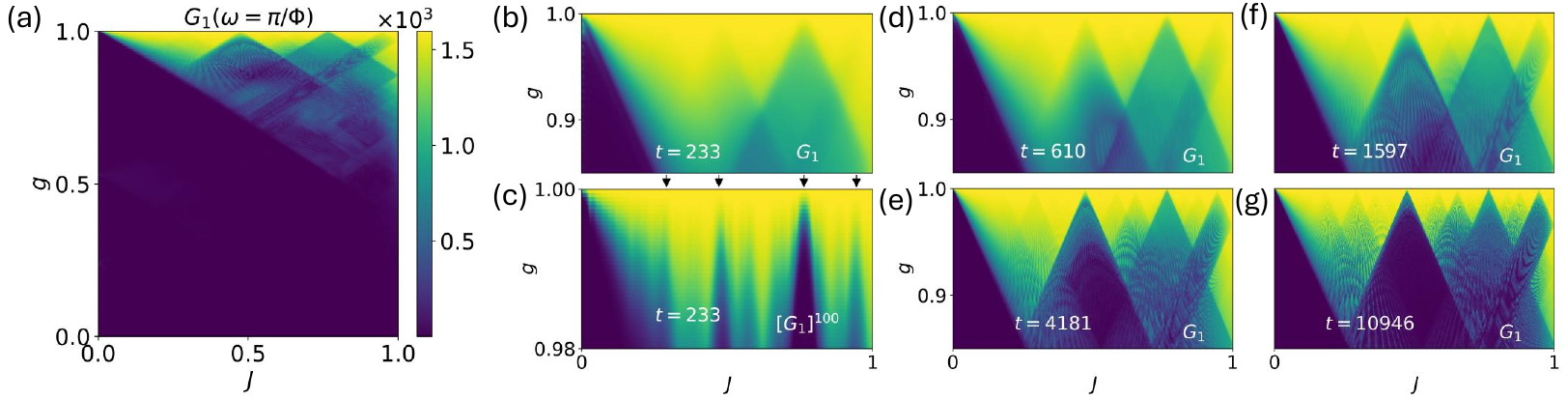}
    \caption{Extended phase diagrams for MGM modes. (a) Dominant Fourier peak of $G_1(\omega=\pi/\Phi)$ in the full $(g,J)$-plane for $t=1597$ time steps. (b), (d)-(e) Zooms for different times $t=233=f_{12},\dots, t=10946=f_{20}$. (c) A large power $k$ of $[G_1(\omega=\pi/\Phi)]^k$, here $k=100$, reveals critical points for small simulation times and system sizes, achievable in current NISQ devices \cite{Mi2021,Mi2022}. 
    Parameters: $N=40$.}    \label{fig:phase_diag_ext}
\end{figure*}

\section{Effective Hamiltonian}
We derive the effective Hamiltonian in Eq.\ (4) of the main text. The effective Hamiltonian in the MZM case holds for $g,J\ll 1$ and can be derived using a Baker-Campbell-Hausdorff (BCH) formula \cite{Dumitrescu2018,Maity2019}. Consider the log of the drive at Fibonacci times
\begin{align}
    H_n=i \ln U_n. 
\end{align}
The deflation rule $U_{n+1}(U_0,U_1)  =  U_n(U_1,U_0U_1)$ implies 
\begin{align}
H_{n+1}(H_0,H_1)  =    	H_{n}\left(H_1,i\ln(e^{-iH_0} e^{-iH_1} )\right).
\end{align}
One can now use BCH to obtain 
\begin{align}
    i\ln(e^{-iH_0} e^{-iH_1})    =    H_0    +    H_1    -\frac{i}{2}[H_0,H_1]+\dots \,.
\end{align}
Repeating, one finds for the quasi-periodic sequence
\begin{align}
    H_n =& f_{n}H_1+f_{n-1}H_0 -\frac{i}{2}((-1)^n+f_{n-2})
    [H_0, H_1]+\dots
\end{align}
Higher orders are given in Ref.\ \cite{Dumitrescu2018}. Neglecting all commutators, the effective Hamiltonian (per time step) is 
\begin{align}
H^\mathrm{MZM}_\mathrm{eff}=\frac{H_n}{f_{n+1}}\simeq 
-\frac{\pi g}{2\Phi }\sum_j X_j-\frac{\pi J}{2 \Phi^2}\sum_j Z_jZ_{j+1}.
\end{align}
This is the transverse field Ising model with renormalized parameters $g\rightarrow g/\Phi $ and $J\rightarrow J/\Phi^2$ and corresponding time evolution operator $U(t)\simeq e^{-iH^\mathrm{MZM}_\mathrm{eff}t}$.

Neglecting all commutators is justified by (i) taking $g,J$ sufficiently small and (ii) observing that the prefactor $-((-1)^n+f_{n-2})/2$ in front of the first commutators grows only linearly in time. This implies that local reorderings in the driving sequence are sufficient to obtain the effective Hamiltonian. 

For the MGM case, we consider small $\delta g =1-g$ and $J$. The transverse field can be written as
$U_1 = P e^{i\pi \frac{\pi \delta g }{2 }\sum_j X_j}$ up to a global phase. The parity operator $P$ commutes with both $U_0$ and $U_1$. Thus, we find for the time evolution operator  $U(t)\simeq P^{N_1(t)} e^{-iH^\mathrm{MGM}_\mathrm{eff}t}$ with effective Hamiltonian
\begin{align}
H^\mathrm{MGM}_\mathrm{eff}\simeq 
-\frac{\pi \delta g}{2\Phi }\sum_j X_j-\frac{\pi J}{2 \Phi^2}\sum_j Z_jZ_{j+1}.
\end{align}

\end{document}